\documentclass{article}

\usepackage[authoryear]{natbib}
\usepackage{mathtools}
\usepackage{float}
\usepackage{amssymb}
\usepackage{pifont}
\newcommand{\xmark}{\ding{55}}%
\usepackage{amsmath}
\usepackage{tabularray}
\usepackage{subcaption}
\usepackage{makecell}
\usepackage{float}
\usepackage[linesnumbered,ruled]{algorithm2e}
\usepackage{PRIMEarxiv}

\usepackage[utf8]{inputenc} 
\usepackage[T1]{fontenc}    
\usepackage{hyperref}       
\usepackage{url}            
\usepackage{booktabs}       
\usepackage{amsfonts}       
\usepackage{nicefrac}       
\usepackage{microtype}      
\usepackage{lipsum}		
\usepackage{graphicx}
\usepackage{doi}
\usepackage{tikz}
\def\checkmark{\tikz\fill[scale=0.4](0,.35) -- (.25,0) -- (1,.7) -- (.25,.15) -- cycle;}
\definecolor{darkgreen}{rgb}{0.0, 0.5, 0.0}
\title{Green Bubbles: A Four-Stage Paradigm for Detection and Propagation
}

\author{
\hspace{1mm}Gian Luca Vriz\\
	Department of Statistics\\
	University of Padova\\
	Via Cesare Battisti, 241, Padova, 35121, Italy \\ \texttt{gianluca.vriz@studenti.unipd.it} \\
	\And
 \hspace{1mm}Luigi Grossi \\
	Department of Engineering and Architecture\\
    University of Parma\\
	 Parco Area delle Scienze, 59, Parma, 43124, Italy \\
	\texttt{luigi.grossi@unipr.it} \\
}

\begin{document}
\maketitle

\begin{abstract}
Climate change has emerged as a significant global concern, attracting increasing attention worldwide. While green bubbles may be examined through a social bubble hypothesis, it is essential not to neglect a \textit{``Climate Minsky''} moment triggered by sudden asset price changes. The significant increase in green investments highlights the urgent need for a comprehensive understanding of these market dynamics. Therefore, the current paper introduces a novel paradigm for studying such phenomena.

Focusing on the renewable energy sector, Statistical Process Control (SPC) methodologies are employed to identify green bubbles within time series data. Furthermore, search volume indexes and social factors are incorporated into established econometric models to reveal potential implications for the financial system.
Inspired by Joseph Schumpeter's perspectives on business cycles, this study recognizes green bubbles as a \textit{``necessary evil''} for facilitating a successful transition towards a more sustainable future. 
\end{abstract}

\keywords{Change Point Detection Model \and Econometrics \and Financial Stability \and Green Bubbles \and Google Trends\footnote{{\textit{\underline{Citation}}: 
Vriz, G. L., Grossi, L.; \textit{Green Bubbles: A Four-Stage Paradigm for Detection and Propagation. arXiv preprint, 2024}.}} 
}

\section{Introduction} \label{introduction}
Since the pioneering research of Nobel Laureate William Nordhaus in the 1970s, scholars have investigated the connections between climate change and the economy. Researchers have conducted several studies to gain a deeper understanding of the correlations between climate change-related risks and asset prices \citep{SALISU2023103383,en14238085,FACCINI2023106948,barnett2020pricing,barnett2023climate}.
As introduced by \cite{giglio2021climate}, these studies gave rise to a new discipline termed \textit{``Climate finance''}, which comprises two primary branches. The first focuses on forecasting climate impacts, while the second examines changes in asset pricing resulting from climate-related events. 

In the prevalent asset pricing theory, the market value of a financial asset reflects the anticipated net present value of its future payoffs. Consequently, the price is conditional on the general expectations that individuals hold for the future, which are driven by both exogenous and endogenous shocks. Global warming can lead to changes in the assets of some of the largest companies resulting in financing problems. On the other hand, the valuation of green companies could see an increasing rise in investment, exceeding their net asset values. As a consequence, a period of financial instability may arise; i.e. when the financial sector is not in perfect sync with the real economy, making the entire system vulnerable to an inevitable \textit{``domino effect''}. \cite{carney2016resolving} outlined how a transition to a low carbon economy may generate a shock of some market values, causing a \textit{``Climate Minsky''} moment. As a result, financial authorities bear the responsibility of monitoring the financial system to detect potential risks on time and determining the necessary actions to prevent adverse consequences. Following the Great Recession, a term known as the \textit{``Green Bubble''} \citep{wimmer2015green} started to gain popularity in academic literature, referring to a situation where the world is over-investing in environmentally-friendly assets. In the last decade, there was a notable boost and subsequent contraction in capital investments within the renewable energy market. From 2005 to 2008, the share of investments going to renewable energy assets was more than tripled before a substantial collapse in the following years \citep{van2022role}.
Notably, the speculative sentiment surrounding investments in clean technology may have been the consequence of an increasing sensibilization on climate change \citep{giorgis2022salvation} and the emergence of new technologies \citep{van2022role}. Thus, studies have revealed a meaningful relationship between clean energy, high technology stocks, and oil prices \citep{kassouri2021oil}. However, the true impact of a new green bubble is still being explored. \cite{ghosh2022deconstruction,ghosh2022bubble} argued that not all financial bubbles are harmful to the economic system, and defined green bubbles under a social bubble hypothesis. They suggested that these bubbles can accelerate investment activities in green energy and support the ecological transition. Nevertheless, these positive outcomes do not come without costs. \cite{borio2023finance} claimed that the risk of a green bubble is not insignificant. Investors have genuine incentives to create bubbles, and private agents are always awaiting some form of public support for eco-investing. However, excessive market behaviour is detrimental not only in terms of financial stability but also to the credibility of the transition process. As highlighted in Schumpeter's work on creative destruction in the 1930s, effective innovation management is crucial for economic development and recovery \citep{Schumpeter2003}. In this view, \cite{cjz2024} discussed how monetary policies could be designed to react to green bubbles.   

\begin{figure}[ht!]
    \centering
    \includegraphics[scale=0.63]{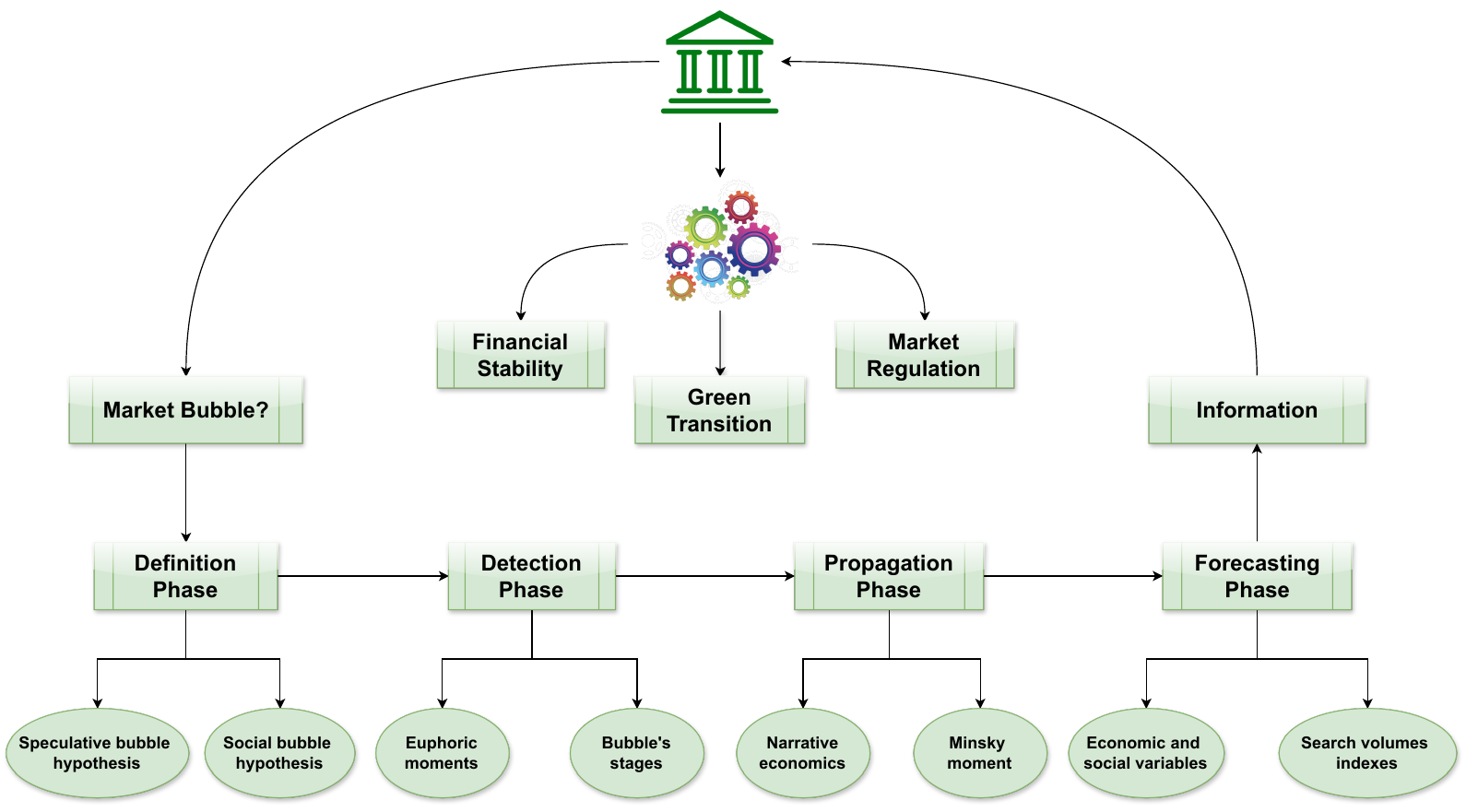}
    \caption{Proposed framework for studying green bubbles.}
    \label{paradigm}
\end{figure}

Recently, \cite{aversa2023scenario} conducted a review using text analytics on the topics of scenario analysis and climate change, emphasizing the need for improvements in the modelling, as well as further research on the formation of possible \textit{``Climate Minsky''} moments. In the strand of this suggestion, the present article tries to develop a new paradigm for detecting and comprehending emerging green bubbles. It highlights the importance of climate change sentiment in predicting future market dynamics, suggesting valuable insights into the potential emergence of \textit{``Climate Minsky''} moments. Figure \ref{paradigm} presents the proposed paradigm, consisting of four main phases that will be illustrated and discussed in the next sections of the paper. The definition phase offers hypotheses beyond the emergence of green bubbles. The detection phase aims to provide suitable tools for examining different stages of bubbles and explosive behaviours within a time series. Moving to the propagation phase, the objective is to investigate potential channels through which social influences may generate green bubbles, leading to potential \textit{``Climate Minsky''} moments. Ultimately, the forecasting phase is designed to predict future market dynamics. Examining trends in keyword search volume and pertinent economic and social data, one gains a deeper understanding of market forces.

This novel approach aims to provide policymakers with a robust framework for ensuring financial stability, while also supporting the shift towards a more ecological economy with novel market regulations.

This is the structure of the paper. Section \ref{literature} details the initial stage of the paradigm, comprising an extensive literature review. Section \ref{detection} examines the detection phase. Section \ref{propagation} delves into exploring the emergence of \textit{``Climate Minsky''} moments, emphasising potential policy implications. Employing an econometric framework, Section \ref{forecasting} is devoted to the final block of the paradigm: the forecasting phase. Lastly, Section \ref{conclusions} brings the analysis to a close, providing a summary of the main findings.

\section{Definition Phase}\label{literature}
The following section is dedicated to exploring multiple interconnected fields.

\subsection{Speculative bubble}
Financial bubbles may be seen as outcomes of market forces. \cite{garber2001famous} characterized bubbles as a vague concept, suggesting that they involve price movements that cannot be fully explained. He also emphasized how comprehending bubbles is a challenging task since they are events that include financial, economic, and psychological factors. The general equilibrium model proposed by \cite{pastor2009technological} predicted that during technological revolutions, stock values of innovative companies may experience bubble patterns. This pattern is reflected in recent boom-bust cycles, such as the dot-com bubble of the early 2000s and the global financial crisis, which may represent two facets of the same underlying phenomenon. The first stage is rooted in technological innovations, while the second cycle is driven by financial innovations. These occurrences are inherent to the progression and assimilation of successive technological revolutions in the market economy. 
Given the challenges of climate change, it is crucial to drive innovation in clean technology. Therefore, public interventions for private green innovation are essential to address global warming. 
\cite{quinn2020boom} proposed a concept called the \textit{``bubble triangle''}, as a warning against excessive market exuberance. Other recent articles, defined green bubbles under a social lens \citep{giorgis2022salvation,ghosh2022bubble,ghosh2022deconstruction}. However, it becomes essential to adopt a critical and rational perspective when assessing the true nature of a green bubble \citep{borio2023finance}.

Joseph A. Schumpeter is known for his theory of innovation and economic growth \citep{andersen2011joseph}. According to Schumpeter, economic growth does not happen linearly, but through waves of economic change driven by technological innovations. Schumpeter has substantially adapted Kondratiev’s wave theory, arguing that scientific and technological progress is the main engine of the economic cycle. From this perspective, new technologies lead to a phase of economic expansion, increasing productivity and economic opportunities. This positive expansion continues to a saturation point, in which new technologies become the standard to follow. Failing to give a boost to the economy, a phase of crisis inexorably starts. For Schumpeter, the economic cycle would be the result of specific market forces that promote innovation, which eventually leads to the development of speculative activities and the need for an inevitable correction cycle. 
\subsection{Bubble detection}
As detailed in the work of \cite{boubaker2022mirror}, methods to identify bubble phases fall into three main categories. The first involves theoretical approaches, which utilize models based on the rational expectations assumption of investors. Alternatively, mathematical procedures employ stochastic analysis to establish a numeric definition of a bubble, often by associating it with deviations from a real martingale process. Empirical methods, on the other hand, encompass hybrid and statistical approaches. In this last category, bubble detection is typically reformulated in terms of explosive behaviours. A large contribution came by \cite{phillips2011explosive,phillips2011dating,phillips2015testing,Phillips2015b,phillips2020real}, which provided a series of papers introducing novel unit root tests for market exuberance. These techniques are widely utilized to detect speculative behaviours in various econometric studies. Quite recently, \cite{monschang2021sup} investigated the performance and suitability of the Supremum Augmented Dickey-Fuller (SADF) test, and its extensions on such a task.
Using Monte Carlo simulations, the majority of the tests display considerable size distortions when the data-generating process is affected by leverage effects. Furthermore, date-stamping results can be highly influenced by the selected data frequency. Despite their important contribution to the econometric literature, statistical tests for explosiveness cannot be universally regarded as a tool for bubble detection \citep{monschang2021sup}. \cite{ENOKSEN2020129} found how bubbles in cryptocurrency markets share common features, such as high volatility regimes, high trading volumes, and high transaction activities. Adopting a fixed-sampled Cumulative Sum control chart (CUSUM), \cite{boubaker2022mirror} proposed a new method for detecting the timing of bubble bursts in global stock markets. 

Drawing from rational expectations theories, bubbles are recognized to have distinct phases: formation, burst, and decline. Nevertheless, a unique definition is lacking in the literature. As previously mentioned, it may depend on a change in the mean of returns, in the variance, in the trading volumes, or associated with explosive behaviours. As said by Nobel laureate Eugene Fama (2016): 
\begin{quote}
``\textit{For bubbles, I want a systematic way of identifying them. It’s a simple proposition. You have to be able to predict that there is some end to it. All the tests people have done trying to do that don’t work. Statistically, people have not come up with ways of identifying bubbles}''.
\end{quote}

Hence, uncovering the fundamental dynamics of a speculative bubble presents a challenge, demanding more comprehensive and unique techniques.

\subsection{Climate Minsky moment}

\cite{carney2016resolving} did not negate the possibility of such momentum during the transition to a low-carbon economy, coning the concept of \textit{``Climate Minsky''} moment. Under specific economic assumptions, \cite{miller2022preventing} outlined how the shift towards a low-carbon economy could potentially lead to the occurrence of \textit{``Climate Minsky''} moments. Such pioneering studies opened the path to a new line of research, in the field of \textit{``Climate Finance''} \citep{giglio2021climate}. Researchers provided evidence on how Google Trends data presents a valid source of information for predicting economic variables and future trading behaviours \citep{preis2013quantifying,financegoogle}. Consequently, search volume indexes may serve as a metric for quantifying investors' sentiment on climate change, being able to anticipate bubble behaviours.

\section{Detection Phase}\label{detection}
This section is devoted to identifying speculative bubbles and euphoric moments in the renewable energy market. To achieve this goal, the study focuses on the monthly data of the global stock index RENIXX (Renewable Energy Industrial Index, ISIN: DE000RENX014) from January 2005 to December 2022. This index is composed of the top thirty renewable energy companies listed on various international stock exchanges and serves as a crucial indicator for investors, offering insights into the overall performance of the market.

\subsection{Tools}
While the SADF and its variants are widely recognized for their efficiency in detecting euphoric moments within a time series, the same cannot be asserted for identifying bubbles' phases \citep{monschang2021sup}. Following the logic of \cite{boubaker2022mirror}, adopting a change point detection strategy might represent a superior option. Employing the data generation process proposed by \cite{phillips2018financial}, a simulation study is carried out to compare the two methodologies mentioned above. The results of the study are available in the Appendix. Based on these findings, the SADF and its variants are utilized to detect euphoric moments, while a change point detection strategy is implemented to identify the bubble's phases.

\subsection{Euphoric moments}
After the global financial crisis, there has been a significant interest in using econometric tests to identify exuberance in asset markets. As a result, new econometric methodologies have been proposed to detect these phenomena. An approach that has gained increasing popularity is the SADF test developed by \cite{phillips2011explosive}, subsequently extended by \cite{phillips2015testing,Phillips2015b} to the Generalized SADF test (GSADF). The primary advantage of employing these techniques is associated with a recursive estimation of ADF tests on sub-samples of the data. Following \cite{vasilopoulos2022exuber}, these tests have their foundation in the following regression equation,
\begin{equation*}
\Delta y_t = a_{r_1,r_2}+\gamma_{r_1,r_2}y_{t-1}+\sum_{j=1}^k \phi_{r_1,r_2}^j
\Delta y_{t-j}+\epsilon_t,
\end{equation*}
where $\{y_t\}_{t=0}^T$ denotes the time series, $\Delta y_{t-j}$ with $j=1,...,k$ are lagged first differences of the series, $\epsilon_t \sim \mathcal{N}(0,\sigma^2_{r_1,r_2})$ are the Gaussian disturbances, and $a_{r_1,r_2},\gamma_{r_1,r_2}$, $\phi^j_{r_1,r_2}$ are regression coefficients. The terms $r_1$ and $r_2$ represent the initial and final points of a sub-sample period. The hypothesis test setup may be summarised as follows,
\begin{equation*}
\left\{\begin{array}{@{}l@{}}
H_0:\gamma_{r_1,r_2}=0,\\
H_1:\gamma_{r_1,r_2}>0,
\end{array}\right.\,
\end{equation*}
with the test statistics for the null hypothesis given by,
\begin{equation}
ADF^{r_2}_{r_1}=\frac{\hat{\gamma}_{r_1,r_2}}{s.e.(\hat{\gamma}_{r_1,r_2})}.  
\label{ADF}
\end{equation}
\cite{phillips2011explosive}, revisited the standard ADF, proposing an approach based on a recursive estimation of \eqref{ADF}. Setting $r_0$ as the minimal sample window
width, the authors defined the supremum of this sequence by,
\begin{equation*}
SADF(r_0)=\sup_{r_2\in[r_0,1]} \left\{ADF_{r_1=0}^{r_2}\right\}.
\end{equation*}
As for the standard ADF, the rejection of the null hypothesis of a unit root necessitates that the SADF statistic surpasses the critical value from its limit distribution in the right tail.
The asymptotic distributions of the SADF test statistics under the null hypothesis are derived by the prototypical model with weak intercept form,
\begin{equation*}
y_t=d T^{-\eta}+\gamma y_{t-1}+\epsilon_t, \quad \epsilon_t \stackrel{\text { i.i.d. }}{\sim}\left(0, \sigma^2\right),
\end{equation*}
with constants $d$ and $\eta$. Under the null hypothesis of $\gamma=1$, the limiting distribution of the test statistics is given by,
\begin{equation*}
\operatorname{SADF}\left(r_0\right) \xrightarrow{d} \sup _{r_2 \in\left[r_0, 1\right]}\left\{\frac{\frac{1}{2} r_2\left[W\left(r_2\right)^2-r_2\right]-\int_0^{r_2} W(r) dr W\left(r_2\right)}{r_2^{\frac{1}{2}}\left\{r_2 \int_0^{r_2} W(r)^2 dr-\left[\int_0^{r_2} W(r) dr\right]^2\right\}^{\frac{1}{2}}}\right\},
\end{equation*}
where $W(\cdot)$ denotes a Wiener process.
The alternative hypothesis of the SADF test allows the occurrence of explosive dynamics in specific segments of the sample.
\cite{phillips2015testing,Phillips2015b} proposed the generalized SADF (GSADF), which covers a larger number of sub-samples compared to the SADF. Setting the minimum window size $r_0$, \eqref{ADF} is estimated for several possible sub-samples, obtained by changing the boundaries, $r_1$ and $r_2$. Formally,
\begin{equation*}
GSADF\left(r_0\right)=\sup _{\substack{r_2 \in\left[r_0, 1\right] \\
r_1 \in\left[0, r_2-r_0\right]}}\left\{\operatorname{ADF}_{r_1}^{r_2}\right\},
\end{equation*}
with its limit distribution under the null hypothesis given by,
\begin{equation*}
GSADF\left(r_0\right) \xrightarrow{d} \sup _{\substack{r_2 \in\left[r_0, 1\right] \\
r_1 \in\left[0, r_2-r_0\right]}}\left\{\frac{\frac{1}{2}r_w\left[W\left(r_2\right)^2-W\left(r_1\right)^2-r_w\right]-\int_{r_1}^{r_2} W(r) d r\left[W\left(r_2\right)-W\left(r_1\right)\right]}{\left(r_w\right)^{\frac{1}{2}}\left[r_w \int_{r_1}^{r_2} W(r)^2 dr-\left(\int_{r_1}^{r_2} W(r) dr\right)^2\right]^{\frac{1}{2}}}\right\},
\end{equation*}
where $r_w=r_2-r_1$. In the SADF, the start and the end of explosive periods can be obtained with the Backward ADF (BADF) test. A similar strategy based on the sequence of Backward SADF (BSADF) statistics was developed in \cite{phillips2015testing,Phillips2015b},
\begin{equation*}
BSADF_{r_2}(r_0)=\sup_{r_1\in[0,r_2-r_0]}\left\{SADF_{r_1}^{r_2}\right\}.
\end{equation*}
Letting $r_e$ and $r_f$ be the start and the end of euphoric moments, an estimate can be obtained as follows.
\begin{gather*}
\hat{r}_e=\inf_{r_2\in[r_0,1]}\big\{r_2:BSADF_{r_2}(r_0)>CI^\alpha_{r_2}\big\},\\ \hat{r}_f=\inf_{r_2\in[\hat{r}_e,1]}\big\{r_2:BSADF_{r_2}(r_0)<CI^\alpha_{r_2}\big\},
\end{gather*}
where $CI^\alpha_{r_2}$ is the critical value of the SADF for the $[r_2T]$ set of observations.
\begin{figure}[htbp!]
    \centering
    \includegraphics[scale=0.535]{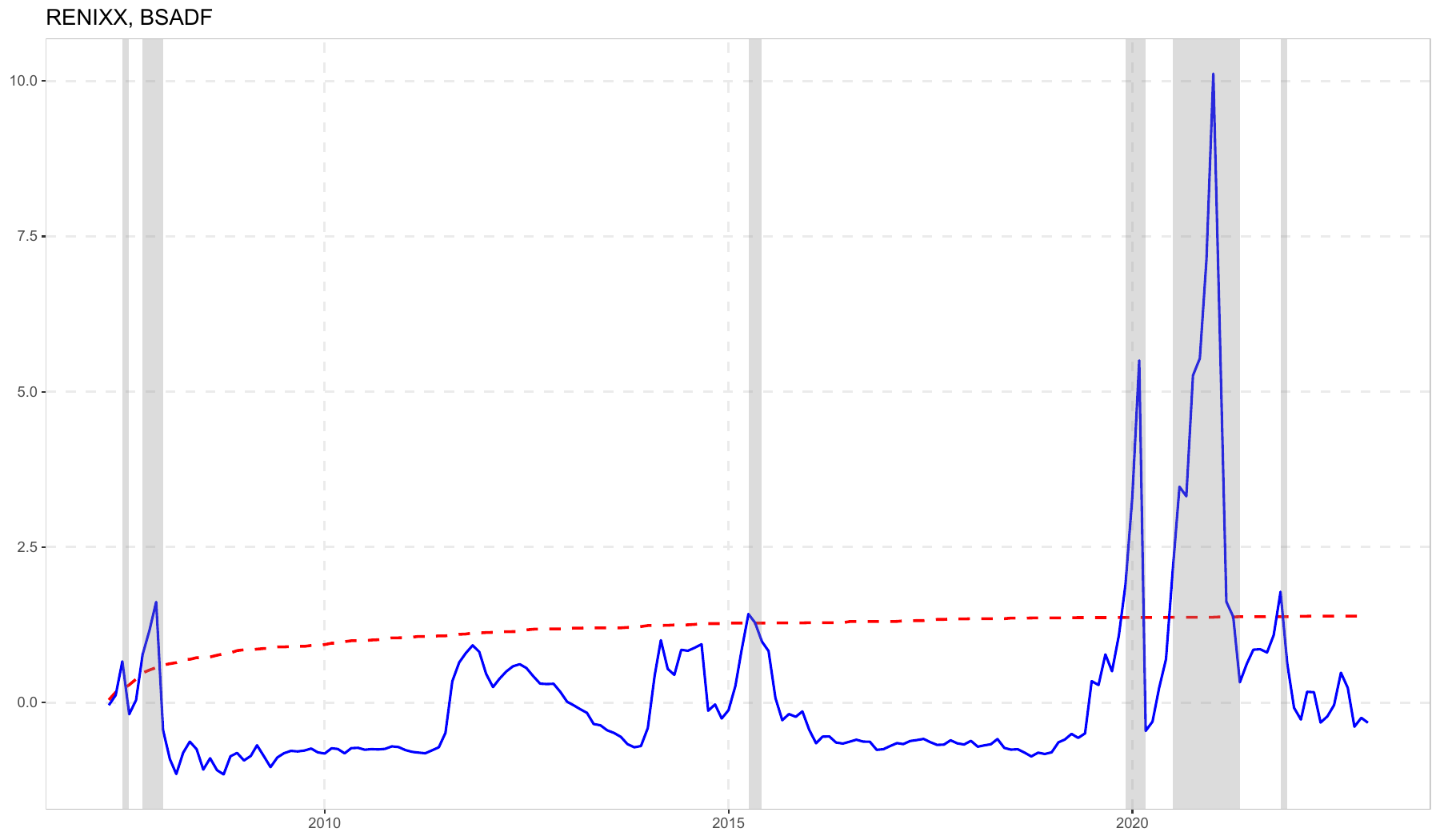}
    \caption{BSADF test statistics with 95\% critical values, RENIXX  index.}
    \label{BSADF}
\end{figure}
Thus, the SADF statistic and its extensions turned out to be reasonable econometric tools for detecting explosive behaviours in time series data\footnote{The BSADF test is implemented using the \texttt{radf} function available in the R package \texttt{exuber}.}. Figure \ref{BSADF} shows the estimated BSADF statistics together with the sequence of 95\% critical values for the RENIXX index. The shaded areas indicate periods in which the statistic is larger than the corresponding critical value, indicating levels of exuberance.
\subsection{Bubbles}
The analysis aims to comprehend the complete bubble process, which includes its formation, burst, and decline stages. The concept of stationarity is fundamental in time series analysis. A bubble can be defined as an exogenous event that induces a remarkable shift in the distribution of a stationary process. Statistically, the log-returns of such a process should exhibit multiple regimes, where breakpoints are associated with the different stages of the bubble's cycle.

Traditional charts, such as CUSUM, rely on a comprehensive understanding of the in-control process; which is not always feasible in practice. Considering this limitation, one may consider utilizing a non-parametric sequential change point model, as outlined in \cite{ross2012two}. Indeed, these models are designed to detect arbitrary alterations in an unknown process distribution. Formally, a sequence of observations $\{x_1,x_2,...\}$ is drawn from the series of random variables $\{X_1, X_2,...\}$ with one or more changes in the distribution at unknown points $\{\tau_1,\tau_2,...,\tau_m\}$. Assuming that the random variables are independent and identically distributed (i.i.d.) inside each segment, the distribution of the sequence can be written as follows,
\begin{equation*}
X_t\sim
\left\{\begin{array}{@{}l@{}}
F_0\quad if \; t\leq\tau_1,\\
F_1\quad if \;\tau_1 < t\leq\tau_2,\\
\quad \quad \quad \quad  \vdots \\
F_{m}\quad if \;  t>\tau_m,\\
\end{array}\right.\,
\end{equation*}
where $F_i$ represents the distribution in each segment. In this regard, the performance of the statistical test widely depends on the assumptions around $F_i$. If there is a fixed sample size\footnote{This is also known as batch change detection scenario \citep{ross2015parametric}.} $\{x_1,x_2,...,x_T\}$, and a change point at some time $\tau$ exists, then observations before this point follow the distribution $F_0$, while the remaining observations follow distribution $F_1$. Testing for a change in the distribution after $k$ observations leads to the following setup of hypothesis,
\begin{equation*}
\left\{\begin{array}{@{}l@{}}
H_0: X_t \sim F_0(x;\theta_0),\quad \forall t,\\
H_1: X_t \sim 
\left\{\begin{array}{@{}l@{}}
F_0(x;\theta_0), \quad t=1,2,...,k,\\
F_1(x;\theta_1), \quad t=k+1,k+2,...,T,
\end{array}\right.\
\end{array}\right.\,
\end{equation*}
where $\theta_i$ represents a vector of unknown parameters of the underlying distribution. This problem can be addressed using a two-sample hypothesis test, in which statistics rely on the assumed distribution. Non-parametric tests may be employed to avoid such assumptions.

\begin{figure}[!ht]
    \centering
    \includegraphics[scale=0.53]{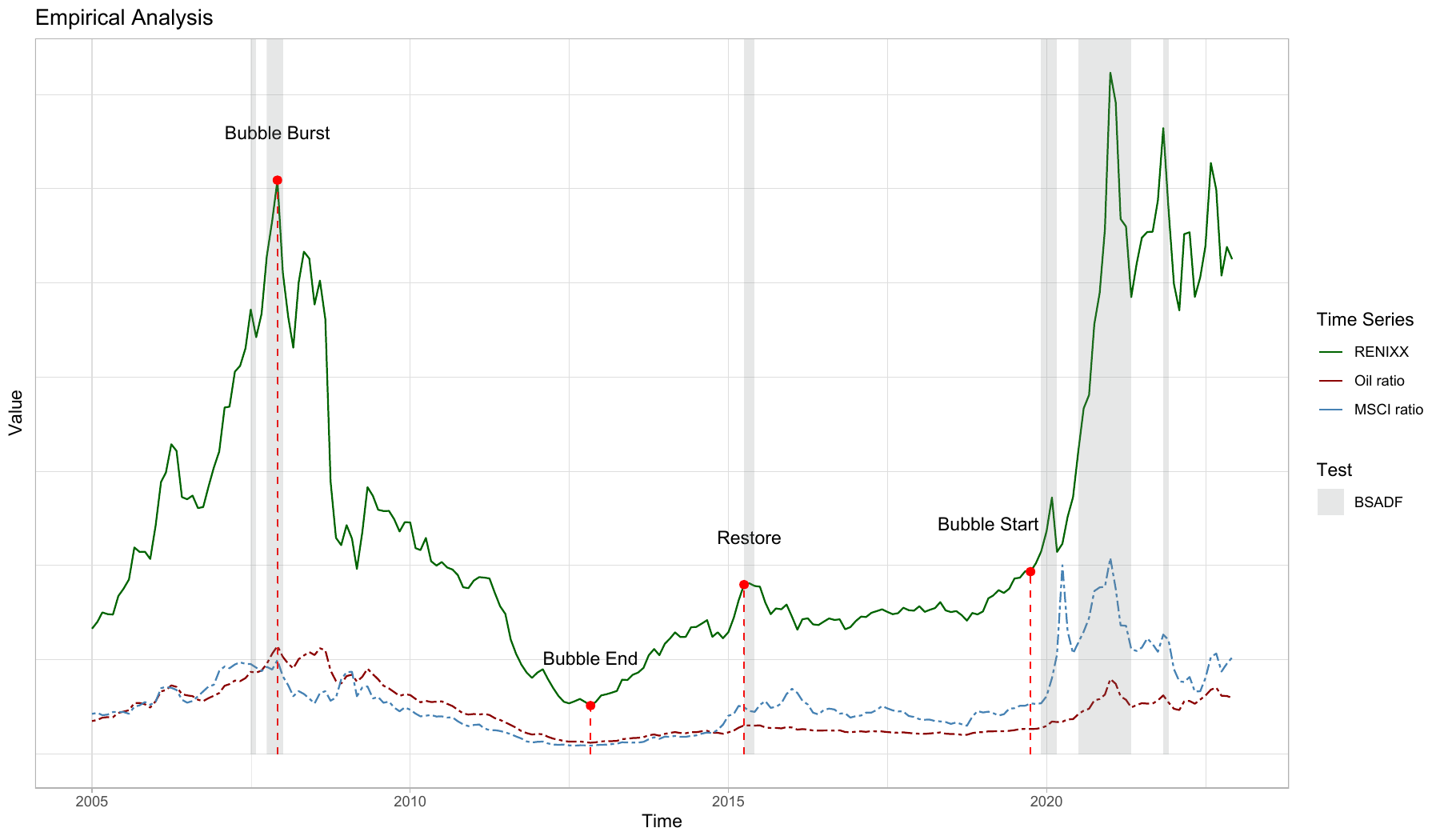}
    \caption{Empirical analysis: RENIXX series in green, oil ratio in dashed red and MSCI ratio in dashed blue. The red points indicate breakpoints detected using the KS-CPM. Dashed areas mark moments of euphoric behaviour identified by the BSADF test statistics.}
    \label{ARL}
\end{figure}

After selecting a two-sample test statistic $D_{k,T}$ and an appropriate threshold point $h_{k,T}$, it becomes possible to test for a change in the distribution of the process immediately following observation $x_k$. However, in practice, the change point $k$ is not known in advance, and therefore $D_{k,T}$ needs to be evaluated for every time point $1<k<T$.
\begin{equation*}
D_T=\max_{k=2,...,T-1}D_{k,T},\\
\hat{\tau}=\arg\max_k D_{k,T},
\end{equation*}
for some threshold $h_T$. Considering a sequential perspective, the above framework can be extended allowing new observations and the presence of multiple change points. Let $x_t$ be the $t^{th}$ observation for $t\in\{1,2,...\}$. When a new observation $x_{t+1}$, is received the sample $\{x_1,x_2,...,x_{t+1}\}$ is treated using a fixed-length perspective and $D_{t+1}$ is computed based on the aforementioned approach. In this sequential setting, the threshold $h_t$ is typically chosen so that the probability of a type I error remains constant over time. Thus, under the null hypothesis,
\begin{equation}
\begin{gathered}
\mathbb{P}(D_1>h_1)=\alpha,\\
\mathbb{P}(D_t>h_t\mid D_{t-1}\leq h_{t-1},...,D_1\leq h_1)=\alpha,\quad t>1.
\end{gathered}
\label{test}
\end{equation}
Generally, the conditional distribution in \eqref{test} is intractable, requiring Monte Carlo simulations to compute $h_t$ for a pre-determined level of $\alpha$ \citep{hawkins2010nonparametric}\footnote{The statistical tests necessary for this purpose can be developed using the R package called \texttt{cpm} \citep{ross2015parametric}. For further details, the reader is invited to consult \cite{ross2012two}.}. Since the Kolmogorov-Smirnov test is widely recognized and commonly utilized for comparing empirical cumulative distribution functions, a sequential Change Point detection Model incorporating the Kolmogorov-Smirnov statistic (KS-CPM) is employed to identify bubble stages. Figure \ref{ARL} shows the number of change points for the log-returns of the RENIXX index\footnote{The model was initialized with a number of observations equal to 20 and an Average Run Length (ARL) of 500. A lower ARL indicates that the control chart is more sensitive and can quickly detect deviations from the norm, while a higher ARL suggests lower sensitivity to changes \citep{ross2015parametric}.}. Additionally, the figure displays the ratios between the RENIXX index and the Morgan Stanley Capital International (MSCI) World Index, as well as the ratio with the Crude Oil West Texas Intermediate Futures. These ratios act as additional indicators of potential speculative behaviours \citep{giorgis2022salvation}, aligning closely with the observed outcomes.

The suggested procedure identifies four change points. The first two are likely connected to the collapse and subsequent conclusion of the documented Clean-Tech bubble \citep{giorgis2022salvation}. The third signifies the restoration of the renewable energy market, while the fourth marks the beginning of a new speculative period, hereafter termed the Climate bubble\footnote{
Given the current global focus on tackling climate change, the recent speculative phase has been labelled the Climate bubble.}. In light of these findings, the RENIXX index may be segmented into three different periods:
\begin{itemize}
    \item Clean-Tech bubble: from January 2005 to November 2012.
    \item No-bubble period: from December 2012 to September 2019.
    \item Climate bubble: from October 2019 to December 2022.
\end{itemize}

\section{Propagation Phase}\label{propagation}
This section serves two primary purposes. Firstly, it aims to examine the commonalities among green bubbles within the renewable energy market. Secondly, it seeks to investigate the potential propagation effects on the stability of the financial system.
\subsection{Dataset}
The dataset used in the subsequent analysis will consist of two principal categories of factors: economic variables, and search volume indexes. Concerning the first group, the following explanatory variables are considered.
\begin{itemize}
    \item Crude Oil West Texas Intermediate (WTI) Futures.
    \item Morgan Stanley Capital International (MSCI) World Index.
    \item Global Economic Uncertainty Policy (EPU) index.
    \item Financial Stress Index (FSI).
    \item Geopolitical Risk (GPR) index\footnote{The WTI Futures and the MSCI World Index are available on the \href{https://www.investing.com/}{Investing.com} platform, while the FSI is sourced following the approach outlined by \cite{FSI}. The EPU index is obtained as proposed by \cite{baker2016measuring} and the GPR as presented in \cite{caldara2022measuring}}. 
\end{itemize}
Regarding the second group of factors, a total of eight search volume indexes have been chosen from Google Trends to measure climate sentiment. Some common terms associated with physical and transition risks are considered:
\begin{itemize}
    \item Climate change:  \textit{``global warming''}, \textit{``natural disasters''}, \textit{``new technology''}, \textit{``carbon price''}, \textit{``carbon tax''} and \textit{``green energy''}.
    \item Green assets: \textit{``energy index''},  \textit{``energy shares''}.
\end{itemize}
To ensure uniformity and consistency, the ``\textit{Glimpse Google Trends Chrome Extension}'' was used to have all indexes in absolute values. The dataset has a total of fourteen variables, covering the period from January 2005 to December 2022.

\subsection{Market drivers}
Bubbles are often linked to the outcomes of the invisible hand. However, investor optimism and herd mentality are recognized as possible main drivers of such events. When investors become excessively optimistic about the future of a specific asset, they initiate a vicious cycle of buying, leading to inflated prices. In 2017, \cite{shiller2017narrative} introduced the concept of \textit{``Narrative Economics''}, to denote the study of how stories may influence economic outcomes. Shiller's proposition seeks to underscore the significant role of narratives in driving economic behaviour, extending beyond the conventional economic models relying on rationality and efficient markets. In essence, the focus lies in the narrative's capacity to impact individuals' beliefs, influencing economic choices. From this perspective, psychological factors, such as the fear of missing out or belief in a new paradigm, may play a relevant role in bubble formation. Therefore, the various phases of the RENIXX index identified in Section \ref{detection} are examined from a climate sentiment standpoint.

\begin{figure}[!ht]
    \centering
     \includegraphics[height=.45\textwidth,width=1.02\textwidth]{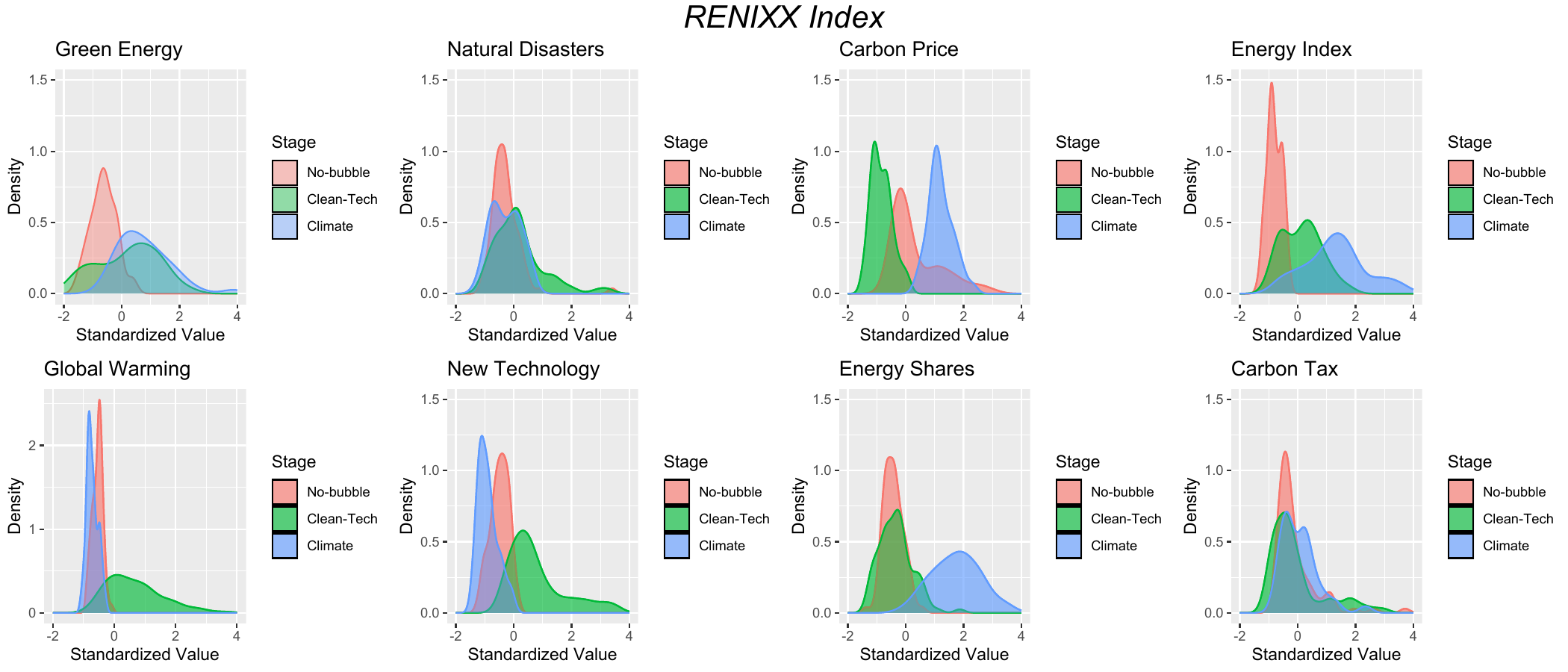}
    \caption{Density analysis, lagged Google Trends $(t-1)$ versus RENIXX index phases $(t)$.}
    \label{descriptive_bubble}
\end{figure}

Figure \ref{descriptive_bubble} depicts the first-lagged series of Google Trends variables\footnote{The series have been standardized to get zero-mean and unit-variance variables. Furthermore, all series are subject to first-order lagging. This means that for the response variable at time $t$, all covariates are lagged at time $t-1$.} related to the three stages identified in the previous section. The initial speculative period is characterized by a high interest in the advent of new technologies, global warming, and new energy sources, i.e. renewable energy. Particularly, the subplot related to the term \textit{``green energy''} delineates how the phase of the Climate bubble in the RENIXX index (depicted in blue) shares similarities with that of the Clean-Tech phase (illustrated in green), differing from the No-bubble phase (represented in red). This indicates a comparable level of research interest during speculative phases for the term \textit{``green energy''}. These results align with the study conducted by \cite{giorgis2022salvation}, highlighting the central narrative of the Clean-Tech bubble as \textit{``salvation and profits''}.
As emphasized by the authors, clean technology not only offered an unprecedented chance for profit, exceeding even that of the Internet but also carried \textit{``quasi-religious''} connotations. Investing in clean tech was viewed as a route to redemption. The narratives surrounding alternative clean and renewable energy technologies, succeeding the anti-nuclear narratives of contamination and apocalypse, fueled the Clean-Tech bubble. Clean and renewable energy evokes spiritual and \textit{``quasi-religious''} imagery of purification, healing, and renewal, in stark contrast to the pervasive imagery of crisis and pollution associated with nuclear energy and fossil fuels.

Apart from the ongoing focus on renewable energies within the Climate bubble phase, the terms \textit{``energy index''} and \textit{``energy shares''}, exhibit higher research volumes compared to previous levels. Perhaps the paradigm proposed by \cite{giorgis2022salvation} requires reconfirmation or even amplification. Following the Paris Agreement, there has been a noticeable interest in climate change across almost all developed nations. This growing interest has been steadily building like a spring over the past years, ready to burst forth, accentuating the \textit{``profit''} dimension of the paradigm. On the contrary, the increasing interest in green policies suggests that individuals have become increasingly concerned about our compulsive dependence on fossil fuels. This, together with a constant interest in green energies, may amplify the quasi-religion concept of \textit{``salvation''}. With these enhancements, there is the potential to generate the virtuous cycles that were absent during the Clean-Tech bubble, establishing new foundations for the ecological transition.

\subsection{Financial stability}
Hyman P. Minsky in his financial instability hypothesis, argued that business cycles naturally emerge within a capitalist economic system \citep{minsky1977financial}. In this view, the economic system and the financial market are strictly connected. A portion of financing may be structured around scheduled payment obligations, with banks playing a central role. Initially, money flows from depositors to banks and then from banks to firms. Subsequently, at later dates, money flows from firms back to banks and from banks to depositors. Therefore, in a capitalist economy, the connections between the past, present, and future are not only defined by capital assets and labour force attributes but also by financial interactions.

From this perspective, not only a rapid rise of green stocks beyond their book value but also a rapid shift towards a low-carbon economy could affect financial stability, triggering a \textit{``Climate Misnky''} moment. This is what \cite{carney2016resolving} referred to as the second climate paradox, i.e. \textit{``success is failure''}. The presence of such a risk is examined within an econometric framework, wherein all variables are transformed by applying the log-difference operator, followed by the application of the first difference operation\footnote{The logarithm transformation was used to make the variance of the variables stationary. The first difference operation was aimed at achieving stationarity in the mean. The logarithm transformation was not applied to the FSI due to the presence of some zeros. In turn, the Breusch-Pagan test was conducted, and it did not reveal any evidence of heteroskedasticity in the time series of first differences.}. Table \ref{stationaritytest} presents the outcomes of both the Augmented Dickey-Fuller (ADF) and Kwiatkowski-Phillips-Schmidt-Shin (KPSS) tests, indicating that all series exhibit $I(1)$ characteristics at a 5\% critical value. Table \ref{granger}  reports the results of the Granger causality test\footnote{The data are on a monthly basis and therefore the Granger causality test was performed considering a lag order of 12.}, as developed by \cite{granger1969investigating}.

\begin{table}[ht!]
\centering
\begin{subtable}[htbp!]{0.45\textwidth}
    \centering
    \begin{tabular}{|c|c|c|}
    \hline
    Variable & ADF & KPSS \\
    \hline
     RENIXX index  &\textcolor{green}{\checkmark}  &\textcolor{green}{\checkmark} \\
     \hline
      MSCI  &\textcolor{green}{\checkmark}  &\textcolor{green}{\checkmark} \\
      \hline
      GPR  &\textcolor{green}{\checkmark}  &\textcolor{green}{\checkmark} \\
      \hline
      FSI &\textcolor{green}{\checkmark}  &\textcolor{green}{\checkmark} \\
      \hline
      EPU &\textcolor{green}{\checkmark}  &\textcolor{green}{\checkmark} \\
      \hline
      Oil price &\textcolor{green}{\checkmark}  &\textcolor{green}{\checkmark} \\
      \hline
      Energy index &\textcolor{green}{\checkmark}  &\textcolor{green}{\checkmark} \\
      \hline
      Energy shares &\textcolor{green}{\checkmark}  &\textcolor{green}{\checkmark} \\
      \hline
      Green energy &\textcolor{green}{\checkmark}  &\textcolor{green}{\checkmark} \\
      \hline
      Global warming &\textcolor{green}{\checkmark}  &\textcolor{green}{\checkmark} \\ 
      \hline
      Natural disasters &\textcolor{green}{\checkmark}  &\textcolor{green}{\checkmark} \\
      \hline
      New Technology &\textcolor{green}{\checkmark}  &\textcolor{green}{\checkmark} \\
      \hline
      Carbon price &\textcolor{green}{\checkmark}  &\textcolor{green}{\checkmark} \\
      \hline
      Carbon Tax &\textcolor{green}{\checkmark}  &\textcolor{green}{\checkmark} \\
      \hline
    \multicolumn{3}{l}{\scriptsize{\textcolor{green}{\checkmark} $p<0.05$}; \textcolor{red}{\xmark} $p>0.05$}
    \end{tabular}
    \caption{Stationarity tests considering a critical value of 5\%.}
\label{stationaritytest}
\end{subtable}
\begin{subtable}[htbp!]{0.45\textwidth}
    \centering
    \begin{tabular}{|c|c|}
    \hline
    \multicolumn{2}{|c|}{\textbf{Granger causality test, FSI }}\\
    \hline
    Variable & P-value\\
    \hline
      RENIXX index & 0.053$^*$\\
     \hline
      Oil price &0.030$^{**}$\\
      \hline
      \multicolumn{2}{|c|}{\textbf{Granger causality test, RENIXX index}}\\
    \hline
    Variable & P-value\\
    \hline
      FSI & 0.435\\
     \hline
      Oil price &0.021$^{**}$\\
      \hline
    \multicolumn{2}{|c|}{\textbf{Granger causality test, oil price}}\\
    \hline
    Variable & P-value\\
    \hline
      RENIXX index & 0.749\\
     \hline
      FSI &0.052$^*$\\
      \hline
      \multicolumn{2}{l}{\scriptsize{$^{***}p<0.01$; $^{**}p<0.05$; $^{*}p<0.1$}}
    \end{tabular}
    \caption{Granger causality tests, \textit{``Climate Minsky''} moments.}
    \label{granger}
    \end{subtable}
    \caption{Econometric analysis, \textit{``Climate Minsky''} moments.}
\end{table}

The analysis delves into the causal connection between the FSI and two variables believed to be potential catalysts of  \textit{``Climate Minsky''} moments: the RENIXX index and the oil price. This essentially refers to a swift surge in the value of green stocks and a rapid transition towards a low-carbon economy.
The results show the existence of causality among the considered variables. The Vector Autoregressive (VAR) model is adopted to investigate how past values of the RENIXX index and the oil price may affect future financial stability conditions. Detailed outcomes are provided in the Appendix.

The findings align with speculative trends. As a bubble emerges, the pressure on the financial system gradually increases until it reaches its breaking point at the bubble burst. This results in a significant decrease in asset returns, influencing financial stability and sparking Minsky moments. 

\section{Forecasting Phase}\label{forecasting}
As outlined in Section \ref{propagation}, forecasting future trends of the renewable energy market could be essential in maintaining financial stability throughout the transition process. Hence, the below section offers reliable prediction strategies for both euphoric moments and forthcoming prices. 

\subsection{Anticipate exuberance}
Euphoric moments are characterized by brief periods of exuberance. These moments are obtained by converting the RENIXX index into a binary variable on the results of the BSADF test. This new variable will take the value of one when an explosive moment is present and zero otherwise.  Previous researchers \citep{agosto2020financial,su2023risks} utilized the logit model to investigate potential influencing factors. 
This study proposes an elastic net regularization technique to enhance model performance. This extension allows an automatic feature selection while simultaneously addressing multicollinearity issues. The elastic-net regularization has the aim to balance the trade-off between $l_1$ and $l_2$ penalties, increasing model stability. Let's assume there are $K=2$ classes, a sample size of length $N$, and a response variable $\{y_i\}_{i=1}^N$. If a number of $p$ features denoted by $\mathbf{x}_i = (x_{i1}, x_{i2}, \ldots, x_{ip})^{\prime}$ for all observations is present, the probabilities of $\{y_i\}_{i=1}^N$ belonging to the interest class may be estimated. For the logistic regression, the probabilities are modelled using the sigmoid function\footnote{The elastic-net regularized logistic regression is implemented using the R package \texttt{caret}.},
\begin{equation*}
      p_i=\mathbb{P}(y_i=1\mid\mathbf{x}_{i})=\frac{1}{1+\exp[-(\beta_0+\boldsymbol{\beta}\textbf{x}_{i})]}, \quad\forall i,
\end{equation*}
where $\boldsymbol{\beta}=(\beta_1,\beta_2,\ldots,\beta_p)$. In this setup, the elastic-net regularization is included in the model by a modification of the objective function,
\begin{equation*}
\begin{gathered}
\arg\min _{\beta_0,\boldsymbol{\beta}} \left\{ - \sum_{i=1}^{N} [y_i\log(p_i)+(1-y_i)\log(1-p_i)] + \lambda \Omega\right\},\\
\Omega=\left[\alpha \|\boldsymbol{\beta}\|_1 +\frac{1}{2}(1-\alpha)\|\boldsymbol{\beta}\|_2^2\right],
\end{gathered}
\end{equation*}
where $\lambda$ is the regularization parameter, while $\|\boldsymbol{\beta}\|_1$ is the $l_1$-norm, $\alpha$ is the mixing parameter, and $\|\boldsymbol{\beta}\|_2$ is the $l_2$-norm. A critical step is the estimate of $\lambda$ and $\alpha$, which are typically chosen by cross-validation. Due to the temporal nature of time series data, performing cross-validation requires special considerations. Therefore, an expanding window cross-validation technique is adopted for tuning both $\lambda$ and $\alpha$\footnote{The expanding window cross-validation was performed starting with an initial set of approximately 120 observations, aligned with the restoring point shown in Figure \ref{ARL}. A fixed window size of 12 months was used, and the \texttt{tuneLength} parameter was set to 15.}. The feature set comprises the first lagged series of both economic variables and Google Trends. To enhance model stability, search volume indexes are standardized, while financial time series are first differentiated before applying standardization. 
Model performance is evaluated by splitting the data into training and test sets\footnote{The test set comprises the last twenty-five observations.}.
The proposed elastic-net regularized logistic regression is compared with both the conventional logit model and the baseline model, which relies exclusively on information from the RENIXX index itself. With the same training and testing set division, the logit model is built using a Backwards Stepwise Selection (BSS) procedure\footnote{The likelihood-ratio test is utilized to compare various nested models, while the Pregibon's link test and the Variance Inflation Factor (VIF) are adopted to assess model's validity. All the results are provided in the online supplementary materials.}. The estimated logit model is composed of four variables, $\Delta_{RENIXX}$, $\Delta_{MSCI}$, and the search volume of the terms \textit{``new technology''} and \textit{``energy index''}. Table \ref{Coef_lasso} shows the results of the considered approaches. 
In both elastic-net regularized logistic regression and the logit model, notable positive effects are observed for $\Delta_{RENIXX}$ and $\Delta_{MSCI}$. Additionally, the search volume of the term \textit{``energy shares''} emerges as another significant variable highlighted by the former model.

\begin{table}[ht!]
    \centering
    \begin{tabular}{|c|c|c|c|}
    \hline
     Variable (t-1)&Elastic-net&Logit&Baseline\\
        \hline
        Intercept & -2.130&-3.600&-2.355\\
        \hline
        Energy index & 0.076&-1.538&/\\
        \hline
        Energy shares &  0.351&/&/\\
        \hline
        Natural disasters & -0.081&/&/\\
        \hline
        New technology &  -0.143&-2.448&/\\
        \hline
        Carbon tax & -0.060&/&/\\
        \hline
         Carbon price &0.080&/&/\\
        \hline
        $\Delta_{RENIXX}$ & 0.611&4.007&1.436\\
        \hline
        $\Delta_{MSCI}$ & 0.141&1.217&/\\
        \hline
        $\Delta_{Oil}$ & 0.077&/&/\\
        \hline
        $\Delta_{EPU}$ & -0.037&/&/\\
        \hline
    \end{tabular}
    \vspace{0.3cm}
    \caption{Variables with their corresponding coefficients for all proposed models.}
    \label{Coef_lasso}
\end{table}
Table \ref{outsampleall} presents the confusion matrices associated with the above-specified approaches for the test set. The elastic-net regularized logistic regression exhibits significantly superior out-of-sample forecasting performance compared with the other counterparts.
Despite an accuracy of 64\%, the logit model also outperforms the baseline model, which shows the poorest performance. These findings emphasize the significance of both economic variables and search volume indexes in predicting future moments of exuberance in the renewable energy sector. This aligns with the concept of  \textit{``Narrative Economics''} developed by \cite{shiller2017narrative}, highlighting the impact of collective beliefs and emotions on financial outcomes.
\begin{table}[ht!]
\centering
\begin{subtable}[!ht]{0.26\textwidth}
\vspace{0.3cm}
    \centering
    \begin{tabular}{|c|c|c|}
    \hline
    N=25&\multicolumn{2}{c}{Reference}\vline\\
    \hline
    Prediction&0&1\\
    \hline
      0 & 35\% &  12\%\\
    \hline
      1 & 65\% & 88\%\\
    \hline
    \multicolumn{3}{|l|}{Accuracy : 0.52}\\
    \hline
    \end{tabular}
    \caption{Confusion matrix, baseline model.}
    \label{outsamplelnaive}
\end{subtable}
\hspace{0.35cm}
\begin{subtable}[!ht]{0.26\textwidth}
    \centering
    \begin{tabular}{|c|c|c|}
    \hline
    N=25&\multicolumn{2}{c}{Reference}\vline\\
    \hline
    Prediction&0&1\\
    \hline
      0 & 47\% & 0\% \\
    \hline
      1 & 53\% & 100\%\\
    \hline
    \multicolumn{3}{|l|}{Accuracy : 0.64}\\
    \hline
    \end{tabular}
    \caption{Confusion matrix, logit model.}
    \label{outsamplelogit}
\end{subtable}
\hspace{0.35cm}
\begin{subtable}[!ht]{0.26\textwidth}
\vspace{0.35cm}
    \centering
    \begin{tabular}{|c|c|c|}
    \hline
    N=25&\multicolumn{2}{c}{Reference}\vline\\
    \hline
    Prediction&0&1\\
    \hline
      0 & 94\% & 25\% \\
    \hline
      1 & 6\% & 75\%\\
    \hline
        \multicolumn{3}{|l|}{Accuracy : 0.88}\\
    \hline
    \end{tabular}
    \caption{Confusion matrix, elastic-net regularized logistic regression.}
    \label{outsample}
\end{subtable}
\caption{Confusion matrix, for the out-of-sample forecasting.}
\label{outsampleall}
\end{table}
\subsection{Short-term forecasts}
The year 2015 marks a significant turning point in global interest regarding climate change. During this year, several significant events and developments contributed to a remarkable growth in public awareness to address the global challenge posed by climate change. Some of the key events include:
\begin{itemize}
    \item Paris Agreement,    \item Clean Energy Investment Forum in Canada,
    \item United Nations Sustainable Development Goals.
\end{itemize}
Furthermore, considering the rising utilization of Google, the potential relationship between search volume indexes and the RENIXX index might be more evident in recent data. Hence, the subsequent analysis focuses solely on observations from January 2015 onwards. 

To develop a robust framework for anticipating future market values, a one-step-ahead forecast exercise on the logarithm transformation of the time series data is conducted. Two baseline models are being considered: a Seasonal Autoregressive Integrated Moving Average (SARIMA) model from a frequentist perspective, and a Bayesian Structural Time Series (BSTS) model without exogenous variables from a Bayesian standpoint. Formally, the structure of a SARIMA(p,d,q)$\times$(P,D,Q)s model is as follows.
\begin{equation}
\phi(L)(1-L)^d\Phi\left(L^s\right)\left(1-L^s\right)^D y_t=\theta(L) \Theta\left(L^s\right) \epsilon_t, \quad \epsilon_t\sim N(0,\sigma^2),
 \label{sarima}
\end{equation}
with,
\begin{equation*}
\begin{gathered}
\phi(L) y_t=\left(1-\phi_1 L^1-\phi_2 L^2-,...,-\phi_p L^p\right) y_t, \\
\theta(L) \epsilon_t=\left(1+\theta_1 L^1+\theta_2 L^2+,...,+\theta_q L^q\right) \epsilon_t,\\
\Phi(L^s) y_t=\left(1-\Phi_1 L^{1s}-\Phi_2 L^{2s}-,...,-\Phi_P L^{Ps}\right) y_t, \\
\Theta(L^s) \epsilon_t=\left(1+\Theta_1 L^{1s}+\Theta_2 L^{2s}+,...,+\Theta_Q L^{Qs}\right) \epsilon_t,
\end{gathered}
\end{equation*}
where $L$ is the lag operator, $y_t$ is the time series at time $t$, and $\epsilon_t$ the error term at time $t$. The BSTS model developed by \cite{scott2014predicting}, is described by the following equations.
\begin{equation*}
\begin{gathered}
 y_t =\underbrace{\mu_t}_{Trend}+\underbrace{\tau_t}_{Seasonality}+\;\;\; \underbrace{\phi y_{t-1}}_{AR(1)}+\;\;\epsilon_t, \quad \epsilon_t\sim\mathcal{N}(0,H_t),\\
\mu_t =\mu_{t-1}+\delta_{t-1}+u_t, \\
\delta_t =\delta_{t-1}+v_t, \\
\tau_t =-\sum_{s=1}^{S-1} \tau_{t-s}+w_t,
\end{gathered}
\end{equation*}
where $\boldsymbol{\gamma}_t = (u_t, v_t, w_t)\sim\mathcal{N}(\boldsymbol{0},\boldsymbol{Q}_t)$ is a vector of independent components of Gaussian random noise. Covariates may be included in \eqref{sarima}, considering an exogenous structure,
\begin{equation}
\phi(L)(1-L)^d\Phi\left(L^s\right)\left(1-L^s\right)^D y_t=\theta(L) \Theta\left(L^s\right) \epsilon_t+\mathbf{X}_{t}\beta,
\label{arimax_eq}
\end{equation}
where $\mathbf{X}_{t}$ is the vector of exogenous variables at time $t$. 
In academic literature, \eqref{arimax_eq} is acknowledged as the SARIMAX(p,d,q)$\times$(P,D,Q)s model\footnote{The R package \texttt{forecast} is utilized to estimate SARIMA models, while the R package \texttt{bsts} is used for the BSTS model.}.

\begin{figure}[ht!]
    \centering
     \includegraphics[scale=0.47]{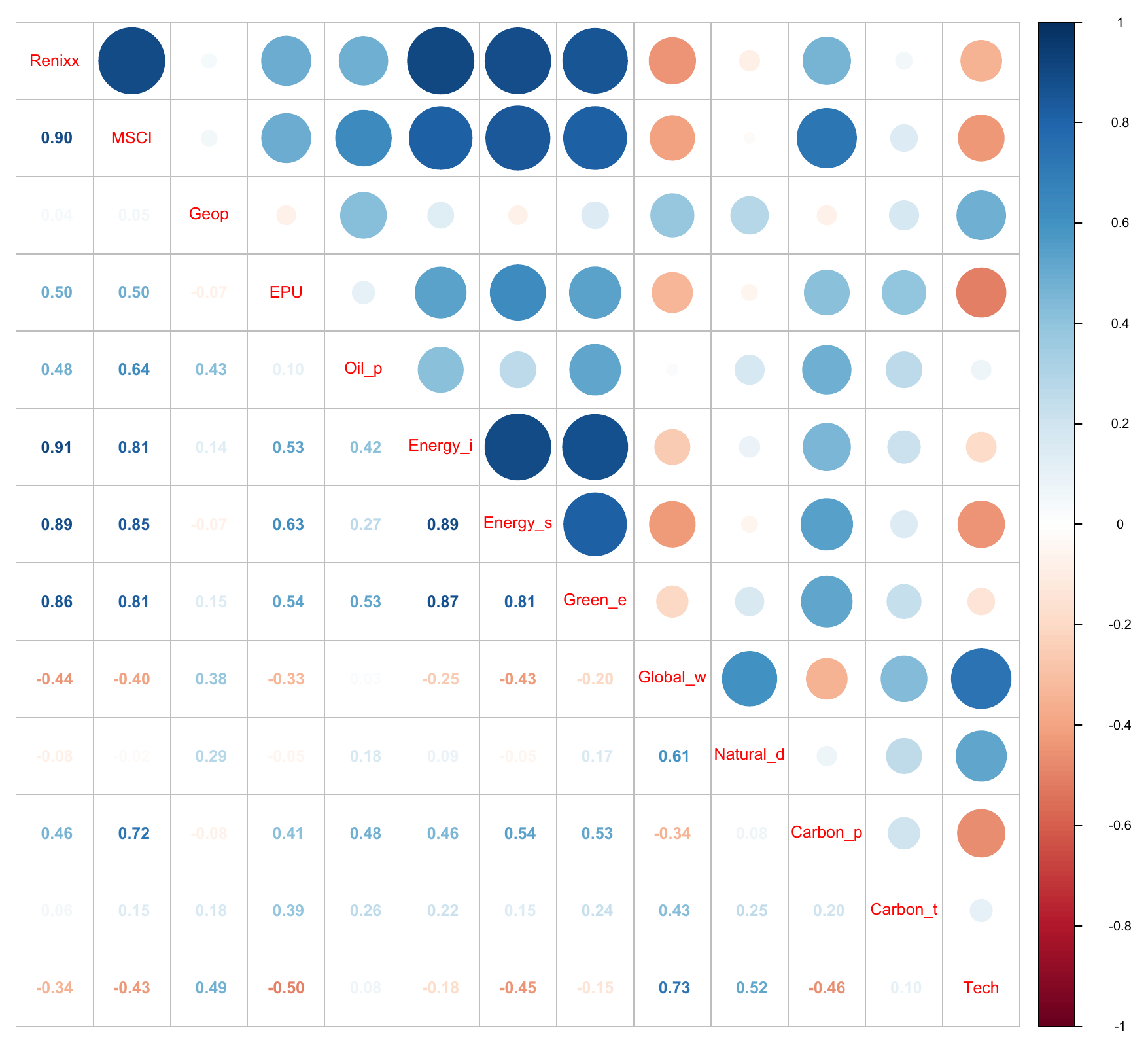}
    \caption{Pearson's correlation coefficient of the RENIXX index with all covariates.}
    \label{corrplot}
\end{figure}

Considering the dataset presented in Section \ref{propagation}, Figure \ref{corrplot} displays Pearson's correlation coefficients among the considered set of covariates. The renewable energy market is part of the overall economy, so the positive correlation with the MSCI should not be surprising. However, there is a notably positive correlation between the RENIXX index and the search volume indexes related to terms,  \textit{``energy index''},  \textit{``energy shares''} and \textit{``green energy''}. So, a SARIMAX(p,q,d)(P,Q,D) model is estimated considering these three covariates as exogenous variables at time $t$. A more robust framework is also employed to prevent overconfidence in the predictions. In other words, the covariate vector in \eqref{arimax_eq} will be fixed at the previous time, $\mathbf{X}_{t-1}$. 
The Granger causality test\footnote{
The p-values presented in Table \ref{granger_renixx} correspond to the respective Wald statistic, calculated using a VAR model of order one.} is used to identify the exogenous variables to include in the model. The results presented in Table \ref{granger_renixx} indicate an unidirectional causality from the search volumes of the term \textit{``energy index''} to the RENIXX index.
\begin{table}[ht!]
    \centering
    \begin{tabular}{|c|c|}
    \hline
    \multicolumn{2}{|c|}{\textbf{Granger causality test, RENIXX index}}\\
    \hline
    Variable & P-value\\
    \hline
      MSCI  &0.136\\
    \hline
      GPR & 0.640\\
      \hline
      EPU  &0.512\\
      \hline
      Oil price &0.632\\
      \hline
      Energy index &<0.01$^{***}$\\
      \hline
      Energy shares &0.507\\
      \hline
      Green energy  &0.240\\
      \hline
      Global warming &0.340\\ 
      \hline
      Natural disasters  &0.340\\
      \hline
      New Technology &0.220\\
      \hline
      Carbon price &0.359 \\
      \hline
      Carbon Tax &0.273\\
      \hline
       \multicolumn{2}{|c|}{\textbf{Granger causality test, energy index}}\\
      \hline
    Variable & P-value\\
    \hline
      RENIXX index&0.131\\
      \hline
       \multicolumn{2}{l}{\scriptsize{$^{***}p<0.01$; $^{**}p<0.05$; $^{*}p<0.1$}}
      \end{tabular}
      \vspace{0.3cm}
    \caption{Granger causality tests, RENIXX index with all covariates.}
    \label{granger_renixx}
\end{table}

Lastly, a VAR model is employed to capture the possible endogenous relationship between the RENIXX index and the term \textit{``energy index''}. Formally, a VAR consists of a set of $K$ endogenous variables $\mathbf{y}_t=(y_{1t},y_{2t},...,y_{Kt})$ for $k=1,...,K$. The VAR(p)-process is therefore defined as,
\begin{equation}
\mathbf{y}_t=\boldsymbol{A}_1\mathbf{y}_{t-1}+,...,+\boldsymbol{A}_p\mathbf{y}_{t-p}+\boldsymbol{\epsilon}_t,
\label{VAR}
\end{equation}
where $\boldsymbol{A}_i$ are $K\times K$ coefficient matrices for $i=1,2,...,p$, and $\boldsymbol{\epsilon}_t$ is a k-dimensional white noise process with $\mathbb{E}(\boldsymbol{\epsilon}_t)=0$ and $\mathbb{E}(\boldsymbol{\epsilon}_t\boldsymbol{\epsilon}_t^T)=\mathbf{\Sigma}_{\boldsymbol{\epsilon}}$. Additional dummy variables are included to account for potential seasonality patterns\footnote{The \texttt{VARselect} function available in the R package \texttt{vars} is adopted to select the lag order of the model, employing the Akaike's Final Prediction Error (FPE) criterion.}. The Johansen procedure \citep{johansen1991estimation} is further developed to test for the existence of cointegration and the suitability of a Vector Error Correction Model (VECM). During the forecasting exercise, \eqref{VAR} may be extended allowing for long-run relationships,
\begin{equation*}
\begin{gathered}
\Delta \mathbf{y}_t=\boldsymbol{\alpha}\boldsymbol{\beta}^{\prime} \mathbf{y}_{t-p}+\boldsymbol{\Gamma}_1 \Delta \mathbf{y}_{t-1}+,...,+\boldsymbol{\Gamma}_{p-1}\mathbf{y}_{t-p+1}+\boldsymbol{\epsilon}_t, \\
\boldsymbol{\Gamma}_i=-(I-\boldsymbol{A}_1-,...,-\boldsymbol{A}_i),\quad i=1,...,p-1,\\
\boldsymbol{\alpha}\boldsymbol{\beta}^{\prime}=-(I-\boldsymbol{A}_1-,...,-\boldsymbol{A}_p).
\end{gathered}
\end{equation*}

The evaluation of all proposed models is based on their performance using future values of the RENIXX index until December 2023. Predictive accuracy is assessed using metrics such as Root Mean Square Error (RMSE) and Symmetric Mean Absolute Percentage Error (SMAPE). The results are presented in Table \ref{forecasting_table}. While the VECM displays the lowest performance, SARIMAX models outperform all competitors. This suggests that simpler exogenous structures offer benefits over complex models, especially considering the limited sample size in the current analysis. The minor differences observed on the used sets of covariates $\mathbf{X}_t$ and $\mathbf{X}_{t-1}$ imply no significant distinction between the two approaches. Nevertheless, the search volume indexes improve the forecasting accuracy of the RENIXX index, particularly when compared to baseline models. These findings justify conducting the Diebold-Mariano Test, whose details are provided in the Appendix.
\begin{table}[ht!]
    \centering
    \begin{tabular}{|c|c|c|}
    \hline
       Model &SMAPE&RMSE \\
       \hline
        SARIMA &0.766&0.066\\
        \hline
        BSTS &0.783&0.070\\
        \hline
        SARIMAX$_{t-1}$&0.674&0.061\\
        \hline
        SARIMAX$_{t}$&0.647&0.060\\
        \hline
        VECM&0.824&0.073\\
        \hline
    \end{tabular}
    \vspace{0.3cm}    \caption{Forecasting accuracy for all proposed models up to December 2023.}
    \label{forecasting_table}
\end{table}

The findings from this analysis align with existing research on the cryptocurrency market, which provides evidence supporting the utility of Google Trends as a valuable resource in financial applications \citep{kristoufek2013bitcoin}. Consequently, policymakers should integrate the dynamics of psychological and sociological factors into their strategies, going beyond conventional economic models.

\subsection{A Schumpeterian perspective}
While certain bubbles can be analyzed through a social hypothesis lens, recent years have been characterized by a significant interest in green investing. As discussed in Section \ref{propagation}, this could represent a new attempt toward the ecological transition. Indeed, climate sentiment plays a crucial role in promoting sustainable and ecological behaviours. Low levels of concern regarding climate issues can undermine efforts to address the global challenge of climate change effectively. At the same time, investors' overestimates should be limited to avoid potential \textit{``Climate Minsky''} moments. 

During the realization of the \textit{``Financial Instability Hypothesis''}, Hyman Minsky was widely influenced by Joseph Schumpeter's ideas on business cycles, and the role of the entrepreneur in economic developments. From this perspective, the recent rise in green investments has similarities with the Railway Mania at the end of the $19^{th}$ century. From Schumpeter's view, railroadization was fueled by technological innovation in the form of steam-powered locomotives and railways. Due to the potential economic benefits, this new technology attracted several investors, leading to a boom in railway investments and disrupting existing transportation methods. Schumpeter coined the term \textit{``creative destruction''} to describe this transformation process, in which old technologies are gradually replaced by new ones. However, the transition did not come without a cost. After the peak of Railway Mania, a bust followed, leading to a significant economic recession. Many railway projects failed, and numerous investors suffered losses. These cycles of boom and bust were named by Schumpeter the \textit{``gales of creative destruction''}.  The history of the Railway Mania is a real example of how periods of rapid technological growth and speculative investments can lead to great innovations and subsequent economic downturns.
Similarly, green bubbles may be seen as a \textit{``necessary evil''} for the success of the green transition. Efficient environmental technologies can be considered innovative activities that create new markets or new forms of competition. Therefore, green bubbles can be described as the financial market's reaction to the notion that environmental sustainability signifies a novel economic frontier, and thus seeks to capitalize on it through speculative transactions. In this sense, green bubbles are a side effect of technological innovations, a source of economic instability. 

\section{Conclusions} \label{conclusions}
The emergence of green bubbles is a recent phenomenon. While they share some common characteristics with speculative activities, their underlying drivers are linked to environmentally sustainable and socially responsible investments. Consequently, it might be appropriate to consider them within the framework of a social bubble hypothesis. However, the likelihood of a \textit{``Climate Minsky''} moment should not be dismissed. The current article aimed to provide a new paradigm to investigate such phenomena.

In Section \ref{detection}, a distinction between euphoric behaviours and the bubble process was outlined, with the introduction of alternative methods for their detection. 
The analysis revealed the existence of explosive moments and identified two distinct bubble stages. Moving on to Section \ref{propagation}, bubbles are revisited under a \textit{``Narrative economics''} lent \citep{shiller2017narrative}. The analysis utilized pre-established climate change-related volume search indexes and found evidence supporting the \textit{``salvation and profits''} paradigm proposed by \cite{giorgis2022salvation}. Moreover,  an econometric framework was adopted to outline the potential of a \textit{``Climate Minsky''} moment, either triggered by a shift in the oil price or renewable energy prices. 
Consequently, Section \ref{forecasting} was devoted to predicting future trends in the RENIXX index. Both economic variables and search volume indexes were found to enhance market predictions. Additionally, the development of logistic regressions outperformed the baseline approach in anticipating future euphoric behaviours. 

As the literature on \textit{``Climate Minsky''} moments is in its early stages, it is crucial to gain a deeper understanding of the possible mechanisms through which changes in asset returns might impact financial stability. 
The current study was conducted from an econometric perspective and presents its limitations for future research.
Firstly, the test adopted to detect bubble behaviours is limited to changes in the process's returns. The methodology may be improved by undertaking a conjoint analysis incorporating additional factors, such as transition volumes. Secondly, the analysis was widely based on Google Trends, which provides insights into popular topics but may not be as comprehensive as data from official journals. Thus, a more extensive textual analysis is needed to have a comprehensive view of the role played by the \textit{``Narrative economy''} on green asset prices. Finally, 
a broader analysis is necessary to assess the impact on each traded company tracked by the RENIXX index.

While Schumpeter discussed various economic phenomena, including business cycles and the role of innovation in economic development, he did not directly focus on speculative bubbles in his writings. However, Schumpeter's perspective suggests that entrepreneurs act as creators of capital while simultaneously disrupting economic equilibrium. From this perspective, green bubbles are both a source of instability and a necessary catalyst to boost the ecological transition.

\section*{Author statement and reproducibility}
Gian Luca Vriz expresses his gratitude for the financial support from the FSE REACT-EU program, Action IV.4 ``\textit{PhD Programs and Research Contracts on Innovation}'' and Action IV.5 ``\textit{PhD Programs on Green Topics}''. Furthermore, the authors reported no potential conflict of interest. Online supplementary materials can be accessed through this \href{https://github.com/GianVriz/Green-bubble-a-novel-paradigm-of-detection-and-propagation}{link}.

\section*{Author contributions}
The initial research idea originated from Gian Luca Vriz, who conducted the analysis, wrote the code using RStudio, and created supplementary material. Luigi Grossi provided conceptual guidance and supported the theoretical framing during the drafting of the paper.

\section*{Acknowledgments}
We express our gratitude to all participants of the conferences where this work was presented. Their constructive feedback has enriched the final draft of this article.
\begin{itemize}
    \item 17th International Conference on Computational and Financial Econometrics (CFE), Berlin, Germany.
    \item 12th International Ruhr Energy Conference (INREC), Essen, Germany.
\end{itemize}

\bibliography{references}
\newpage 

\section*{Appendix}
\setcounter{figure}{0}
\setcounter{table}{0}
\renewcommand{\thefigure}{A.\arabic{figure}}
\renewcommand{\thetable}{A.\arabic{table}}
The following appendix offers details regarding the analysis presented within the article.
\subsection*{Simulations}
To assess the effectiveness of the KS-CPM described in Section \ref{detection} alongside the BSADF test, a modified version of the data generation process presented in \cite{phillips2018financial} is utilized. Formally,
\begin{equation*}
y_t=
\left\{\begin{array}{@{}l@{}}
aT^{-\eta}+y_{t-1}+\epsilon_t, \quad t\in N_0 \cup N_1,\\
\delta_T y_{t-1}+\epsilon_t,\quad t\in B,\\
\gamma_T y_{t-1}+\epsilon_t,\quad t\in C,
\end{array}\right.\,
\end{equation*}
where $\epsilon_t \sim$ EGARCH(1,1), $\delta_T=1+c_1T^{-\alpha}$, and $\gamma_T=1-c_2T^{-\beta}$.

This configuration facilitates the incorporation of leverage effects, which are observed in financial variables. The set $B$ covers the period in which the bubble inflates, the set $C$ represents the period of bubble collapse, and the set $N_0 \cup N_1$ denotes the no-bubble periods situated at the beginning and end of the sample, respectively. Additionally, $T=y_0=100$, $a=c_1=c_2=1$, while the parameters $\alpha,\beta,\eta,B,C$ vary to consider three alternatives collapse patterns. Within this framework, a comparison is drawn between the KS-CPM and the BSADF test. 

For each collapse pattern, one thousand Monte Carlo simulations are performed, and the outcomes are displayed in Table \ref{results_ks}. 
Model precision is evaluated based on how accurately the proposed methodologies identify the bubble throughout its three distinct phases: formation, burst, and decline. Additionally, the RMSE metric is applied within the remaining sub-sample of simulations to assess accuracy by comparing the actual and reported timing of each bubble phase.
\begin{equation*}
{\displaystyle \operatorname {RMSE} ={\sqrt {{\frac {1}{3n}}\sum _{i=1}^{n}\sum _{j=1}^{3}(\widehat{X}_{i,j}-X_{i,j})^{2}}}},
\end{equation*}

where $\widehat{X}_{i,j}$ represents the estimated time of the $j$-th phase within the $i$-th simulation, while $X_{i,j}$ denotes the corresponding real value.
\begin{table}[htbp!]
    \centering
    \begin{subtable}[htbp!]{0.25\textwidth}
    \centering
    \begin{tabular}{|c|c|c|}
    \hline
     \textbf{Sudden}&Correctness&RMSE\\
       \hline
        BSADF &54.3\%&3.165\\
        \hline
        KS-CPM&84.4\%&2.190\\
        \hline
    \end{tabular}
    \caption{One thousand Monte Carlo simulations, sudden collapse.}
\end{subtable}
\hspace{1.3cm}
    \begin{subtable}[htbp!]{0.25\textwidth}
    \centering
    \begin{tabular}{|c|c|c|}
    \hline
    \textbf{Disturbing}&Correctness &RMSE\\
       \hline
        BSADF &84.3\%&4.678\\
        \hline
        KS-CPM&91.4\%&1.761\\
        \hline
    \end{tabular}
    \caption{One thousand Monte Carlo simulations, disturbing collapse.}
\end{subtable}
\hspace{1.6cm}
    \begin{subtable}[htbp!]{0.25\textwidth}
    \centering
    \begin{tabular}{|c|c|c|}
    \hline
        \textbf{Smooth}&Correctness &RMSE\\
       \hline
        BSADF &81.6\%&6.394\\
        \hline
        KS-CPM&89.4\%&2.067\\
        \hline
    \end{tabular}
    \caption{One thousand Monte Carlo simulations, smooth collapse.}
\end{subtable}
\vspace{0.4cm}
    \caption{One thousand Monte Carlo simulations, bubble patterns.}
    \label{results_ks}
\end{table}

\subsection*{Financial ramifications, VAR}
Table \ref{VAR_OFR} presents the results of the estimated VAR(1) model used to examine the impact of the RENIXX index and oil prices on future financial stability conditions.
\begin{figure}
    \centering
 \includegraphics[scale=0.55]{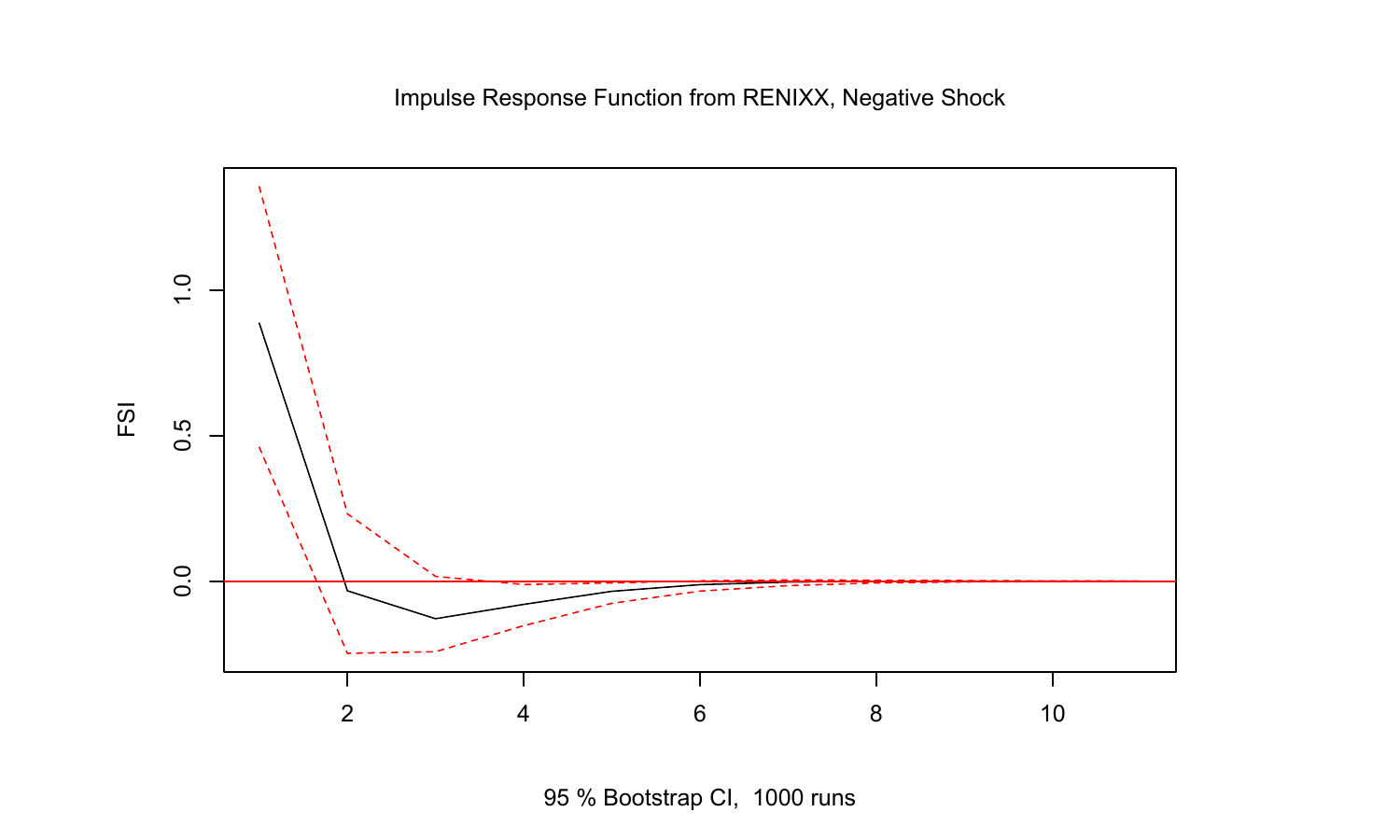}
    \caption{Impulse response function of the FSI from a negative shock in the RENIXX index.}
    \label{Negative_shock}
\end{figure}
\begin{table}
\centering
\resizebox{0.6\textwidth}{!}{%
\begin{tabular}{|r|r|r|r|r|}
\hline
\multicolumn{5}{|c|}{\textbf{FSI}}\\
  \hline
Variable & Estimate & Std. Error & t value & Pr($>$$|$t$|$) \\ 
  \hline
Renixx.l1 & 4.061 & 1.415 & 2.870 & > 0.01$^{***}$ \\ 
  FSI.l1 & 0.366 & 0.083 & 4.433 & > 0.01$^{***}$ \\ 
  Oil\_p.l1 & 0.352 & 1.060 & 0.332 & 0.740 \\ 
  sd1 & 0.448 & 0.521 & 0.860 & 0.391 \\ 
  sd2 & 0.848 & 0.515 & 1.647 & 0.101 \\ 
  sd3 & -0.275 & 0.521 & -0.528 & 0.598 \\ 
  sd4 & 0.279 & 0.513 & 0.543 & 0.588 \\ 
  sd5 & 0.483 & 0.518 & 0.932 & 0.352 \\ 
  sd6 & 0.135 & 0.516 & 0.262 & 0.794 \\ 
  sd7 & 0.634 & 0.513 & 1.236 & 0.218 \\ 
  sd8 & 0.617 & 0.513 & 1.203 & 0.230 \\ 
  sd9 & 0.848 & 0.514 & 1.650 & 0.101 \\ 
  sd10 & 0.160 & 0.517 & 0.310 & 0.757 \\ 
  sd11 & 0.272 & 0.512 & 0.531 & 0.596 \\ 
   \hline
\multicolumn{5}{|c|}{\textbf{RENIXX index}}\\
  \hline
Variable & Estimate & Std. Error & t value & Pr($>$$|$t$|$) \\ 
  \hline
Renixx.l1 & 0.265 & 0.080 & 3.322 & > 0.01$^{***}$ \\ 
  FSI.l1 & -0.003 & 0.005 & -0.727 & 0.468 \\ 
  Oil\_p.l1 & 0.050 & 0.060 & 0.837 & 0.403 \\ 
  sd1 & -0.039 & 0.029 & -1.311 & 0.191 \\ 
  sd2 & -0.061 & 0.029 & -2.101 & 0.037$^{**}$ \\ 
  sd3 & -0.004 & 0.029 & -0.133 & 0.894 \\ 
  sd4 & -0.058 & 0.029 & -2.004 & 0.046$^{**}$ \\ 
  sd5 & -0.046 & 0.029 & -1.572 & 0.117 \\ 
  sd6 & -0.026 & 0.029 & -0.900 & 0.369 \\ 
  sd7 & -0.041 & 0.029 & -1.424 & 0.156 \\ 
  sd8 & -0.043 & 0.029 & -1.500 & 0.135 \\ 
  sd9 & -0.067 & 0.029 & -2.326 & 0.021$^{**}$ \\ 
  sd10 & -0.035 & 0.029 & -1.186 & 0.237 \\ 
  sd11 & -0.038 & 0.029 & -1.304 & 0.194 \\ 
   \hline
      \multicolumn{5}{|c|}{\textbf{Oil price}}\\
      \hline
Variable & Estimate & Std. Error & t value & Pr($>$$|$t$|$) \\ 
  \hline
Renixx.l1 & -0.096 & 0.096 & -1.000 & 0.319 \\ 
  FSI.l1 & --0.023 & 0.006 & -4.110 & > 0.01 $^{***}$\\ 
  Oil\_p.l1 & 0.169 & 0.072 & 2.340 & 0.020 $^{**}$\\ 
  sd1 & 0.000 & 0.035 & 0.005 & 0.996 \\ 
  sd2 & 0.023 & 0.035 & 0.655 & 0.513 \\ 
  sd3 & 0.002 & 0.035 & 0.048 & 0.962 \\ 
  sd4 & 0.029 & 0.035 & 0.841 & 0.401 \\ 
  sd5 & 0.011 & 0.035 & 0.300 & 0.764 \\ 
  sd6 & -0.005 & 0.035 & -0.149 & 0.881 \\ 
  sd7 & -0.036 & 0.035 & -1.033 & 0.303 \\ 
  sd8 & -0.005 & 0.035 & -0.154 & 0.877 \\ 
  sd9 & -0.004 & 0.035 & -0.105 & 0.917 \\ 
  sd10 & -0.025 & 0.035 & -0.713 & 0.477 \\ 
  sd11 & -0.037 & 0.035 & -1.073 & 0.285 \\ 
   \hline 
   \multicolumn{5}{l}{\scriptsize{$^{***}p<0.01$; $^{**}p<0.05$; $^{*}p<0.1$}}
\end{tabular}
}
\vspace{0.3cm}
\caption{Financial stability, VAR(1) model results.}
    \label{VAR_OFR}
\end{table}
The results indicate a significant causality between the RENIXX index at the previous date and the FSI, while the coefficient of the first-lag oil price is not statistically significant. To delve deeper, Figure \ref{Negative_shock} shows the impulse response function from a one-standard-deviation negative shock in the RENIXX index. 
A positive coefficient in the VAR(1) model indicates a positive linear relationship between the RENIXX index and the FSI, which means that past values of the predictor variable have a positive impact on the current value of the response variable. On the other hand, the impulse response function underlines an inverse effect immediately following a shock. Put differently, there is a short-term negative deviation from the expected positive relationship between the variables. As the system stabilizes, the long-term positive relationship reasserts itself. The findings are consistent with speculative patterns. When a bubble forms, there is a gradual increase in pressure on the financial system, peaking at the bubble burst. This leads to a drastic decrease in asset returns, affecting financial stability and potentially triggering a Minsky moment.

When examining the response to a specific shock, the standard impulse response function assumes that all other variables in the system remain constant. While this is a commonly used approach, it may be too restrictive in practice. Additionally, it is essential to extend the VAR(1) model to properly satisfy validation tests, such as the multivariate Portmanteau test for autocorrelation, the multivariate ARCH test for heteroskedasticity, and the multivariate Jarque-Bera test for normality \citep{pfaff2008var}. Therefore, further developments are necessary to have more robust results. Nevertheless, the current analysis has provided valuable insights into the financial consequences of a sudden and unexpected negative shock in the renewable energy sector. In contrast, similar evidence is not found for the oil price, which has an insignificant coefficient.
\subsection*{Robustness analysis}
The Diebold-Mariano (DM) test is a statistical test used to compare the forecast accuracy of two competing models \citep{diebold2002comparing}. It evaluates whether one model significantly outperforms the other in terms of forecasting accuracy. Given two models and a series of loss differential $\{d_t\}_{t=1}^T$, the DM test statistics is computed as follows,
\begin{equation}
DM = \frac{\bar{d}-\mu}{\sqrt{\frac{s_{d}^2}{T}}} \xrightarrow{d}\mathcal{N}(0,1),
\label{DM}
\end{equation}
where $\bar{d}=\sum_{t=1}^T d_t/T$ represents the sample mean loss differential with mean $\mu$, while $\sqrt{s_d^2/T}$ is a consistent estimate of the standard deviation of $\bar{d}$, and $T$ stands for the total number of comparisons. As reported in \eqref{DM}, the test assesses whether the differences in forecast errors are statistically significant, providing insights into whether one model is superior for prediction purposes. 
Table \ref{DM_table} presents the outcomes of the analysis, which compares the baseline models to the SARIMAX models examined in Section \ref{forecasting}. In both scenarios, there is a statistically significant difference concerning baseline models. These results provide additional support for the findings discussed in Section \ref{forecasting}.
\begin{table}[!ht]
    \centering
    \begin{tabular}{|c|c|c|}
    \hline
    \multicolumn{3}{|c|}{\textbf{One-step ahead forecasts}}\\
    \hline
    Model & P-value &P-value\\
        \hline
      SARIMAX$_{t-1}$ & \textcolor{darkgreen}{$Benchmark$}&0.403\\
      \hline
       SARIMAX$_{t}$ & 0.597 & \textcolor{darkgreen}{$Benchmark$}\\
    \hline
      SARIMA & 0.069$^*$&0.040$^{**}$\\
     \hline
      BSTS & 0.093$^{*}$&0.021$^{**}$\\
      \hline
      \multicolumn{2}{l}{\scriptsize{$^{***}p<0.01$; $^{**}p<0.05$; $^{*}p<0.1$}}
    \end{tabular}
    \vspace{0.3cm}
    \caption{Diebold-Mariano test. The alternative hypothesis indicates superior performance in the benchmark.}
    \label{DM_table}
    \end{table}

\end{document}